\documentclass[nofootinbib,prb,twocolumn,amsmath,amssymb,aps,]{revtex4-1}
\usepackage[utf8]{inputenc} 
\usepackage{feynmp}
\usepackage{graphicx}
\usepackage{adjustbox}
\usepackage{dcolumn}
\usepackage{bm}
\usepackage{caption}
\usepackage{subcaption}
\usepackage{comment}
\usepackage{hyperref}
\usepackage{feynmp}
\usepackage{float}
\hypersetup{
     colorlinks   = true,
     linkcolor    = blue,
     citecolor    = blue
}
\usepackage[all]{hypcap}
\newcommand{\expect}[1]{\langle {#1} \rangle}

\newcommand{\sign}{\text{sign}}

\begin{document}
\preprint{APS/123-QED}

\title{Role of boundary conditions, topology and disorder in the chiral magnetic effect in Weyl semimetals}

\author{Yahya Alavirad$^1$}
\author{Jay D. Sau$^1$}

\affiliation{
$^1$Department of Physics, Condensed Matter theory center and the Joint Quantum Institute, University of Maryland, College Park, MD 20742.
}

\date{\today}

\begin{abstract}
Quantum field theory predicts Weyl semimetals to possess a peculiar response of the longitudinal current density to the application of a 
DC magnetic field. Such a response function has been shown to be at odds with a general result showing the vanishing of the bulk 
current in an equilibrium system on any real material with a lattice in an external magnetic field. Here we resolve this apparent contradiction by 
introducing a model where a current flows in response to a magnetic field even without Weyl nodes. 
We point out that the previous derivation of a vanishing CME in the limit of vanishing real frequency is a consequence of the assumption of periodic boundary conditions of the system. A more realistic 
system with open boundary conditions is not subject to these constraints and can have a non-vanishing CME.
Consistent with recent work, we found the finite frequency CME to be 
non-vanishing in general when there was a non-vanishing Berry curvature on the Fermi surface. This does not necessitate having a 
topological Berry flux as in the case of a Weyl node.
Finally, we study how the perturbation theory in magnetic field might be more stable in the presence of disorder.
Using the standard diagrammatic treatment of 
disorder within the Born approximation, we have found that in a realistic disordered system, the chiral magnetic 
response is really a dynamical phenomena and vanishes in the DC limit. 
\end{abstract}

\maketitle

\section{Introduction}
Weyl semimetals, which are three-dimensional analogues of graphene, have generated a lot of interest in recent years because of 
the combination of their peculiar properties  \cite{Burkov2015,Hosur2013a,Xu2011a,Wan2011,Sekine2015} and experimental accessibility~\cite{Lv2015a,Lu2015a,Liu2014,Borisenko2014,Xu2015a}.
Unlike graphene, the gapless nature of the Weyl points in the energy spectrum of a Weyl semimetal are protected by topology through 
the presence of a non-zero Berry flux in momentum space\cite{Wan2011}. The non-zero Berry flux has certain unique characteristics such as chiral Landau 
levels when subjected to a magnetic field~\cite{Potter2014,Haldane2014}. Electrons in the zero energy Landau levels in a Weyl semimetal propagate either parallel or antiparallel 
to the magnetic field and can form a closed loop only with the aid of Fermi arc states on the surface of the Weyl semimetals \cite{Wan2011}. 
Recently some  evidence for such Fermi arcs \cite{Xu2011} and the chiral Landau levels \cite{Jeon2014} has become available.
However, the Landau level trajectories of electrons by themselves do not form a macroscopic response function that can be measured without 
direct reference to the single electrons.
On the other hand, the topological Berry flux in Weyl semimetals is also predicted to give rise to a such a response through the so-called 
 "chiral anomaly" in 
three dimensions known from quantum field theory\cite{Hosur2013a,Nielsen1983}. It has been shown that this chiral anomaly could be applicable to 
Weyl semi-metals in the solid state
solid state systems
in the form of the "chiral magnetic effect" (CME)  
\cite{Vazifeh2013,Goswami2013,Zhou2013,Chang2015,Chen2013,Basar2014,Landsteiner2014,Zyuzin2012,Grushin2012}. 

The CME, which is originally a prediction from the continuum field theory of Weyl Fermions in three dimensions, has been the subject of some debate 
when applied to solid state systems on a lattice. Lattice regularization itself is known to limit Weyl points to exist in pairs so as to ensure 
the vanishing of the total Berry flux in momentum space. Denoting the separation in energy of a pair of such Weyl points by $\delta k_0$,
the CME predicts a current $\mathbf{j}=\Big(\frac{e}{2\pi}\Big)^2\delta k_0 \mathbf{B} $ in response to the application of a magnetic field 
$\mathbf{B}$. This is a rather unusual prediction since in the solid state, with the exception of superconductors, the flow of a current always 
requires an applied electric field. The subtle nature of the field theory prediction was further substantiated by the demonstration of  regularization schemes
where the CME would not occur in Weyl semimetals \cite{Basar2014,Landsteiner2014}.
Using different limits from field theory, a variety of other conclusions were reached for the existence of the CME, such as a critical momentum space
 separation of the Weyl points \cite{Grushin2012}, presence of a gap \cite{Zyuzin2012}.  
Semiclassical analysis\cite{Zhou2013,Chang2015} of the magnetic field response also concluded the CME to be absent in Weyl semimetals. 
Following this, direct (numerical) linear response calculations  of CME for specific lattice models  \cite{Goswami2013,Chen2013} of Weyl semimetal it is concluded  
that the CME can indeed occur as predicted by field theory in the appropriate momentum and frequency limit. 
However, the numerical confirmation of the CME by linear response studies of lattice models does not address the counter-intuitive nature of the 
CME i.e. how a current can flow in response to just a magnetic field. In fact, Vazifeh and Franz ~\cite{Vazifeh2013} and later Yamamoto\cite{Yamamoto2015} have shown rigorously that the 
current in  thermal equilibrium  in any solid state material must vanish in the absence of an electric field.

 In this paper, we address these questions by studying the magnetic field response 
of the current in metals in different situations. We start in Sec.~\ref{II} by using a model Hamiltonian to demonstrate 
that an adiabatically increasing magnetic field can generate a charge current  along the direction of the magnetic field even without 
any topological properties such as Weyl nodes in the dispersion. This establishes that not only is a CME-like current response possible, it is 
not unique to topological systems. In Sec.~\ref{III} we carefully re-examine the linear response properties and distinguish two kinds of linear response 
namely - thermal equilibrium response and dynamical response in the DC limit.  In Sec.~\ref{IIIA}, we review how the equilibrium linear response must 
identically vanish. Furthermore, we show that for finite wave-vector magnetic fields in periodic boundary condition systems the DC limit of the dynamical 
response coincides with the equilibrium response and therefore also vanishes. In Sec.~\ref{IIIC}, we show that the linear response is not as straightforward 
in systems with open boundary conditions where a uniform magnetic field may be applied. In this case, the DC limit of the dynamical response differs 
from the vanishing equilibrium response and remains finite. Finally in Sec.~\ref{IV}, we show that while disorder might be used to make the notion of a perturbative
magnetic field more well-defined, it still leads to a vanishing CME response due to scattering.

\section{Chiral magnetic response of conventional systems at finiite $B$ field}\label{II}
In this section we present an example of a system which develops a DC current in response to the application 
of a DC magnetic field $\mathbf{B}$ that is parallel to the direction of the current. Therefore, in a sense we will see 
that the key surprising aspect of the chiral magnetic response i.e. a current response to a magnetic field is not
only possible, but is not unique to non-topological systems.

The model we study  is described by the  Hamiltonian 
\begin{equation}\label{24}
\hat{H}(\mathbf{k})=\mathbf{k}^4+\alpha k_F k_z^3 -k_F ^4 
\end{equation}
which is parametrized by  $k_F$ and $\alpha$. In the limit $\alpha\rightarrow 0$, $k_F$ describes the Fermi wave-vector 
of the system. The parameter $\alpha$ is key to breaking time-reversal  and inversion symmetries along the $z$-direction, which 
are symmetries that would forbid a current response. We will choose this parameter to be small i.e. $\alpha\ll 1$, so 
that the modification of the dispersion around the Fermi surface can be computed perturbatively in $\alpha$.

 Applying a constant magnetic field along the $z$ axis in Landau gauge changes it to 
\begin{align}\label{25}
&\hat{H}(k_x-eBy,-i\partial_y,k_z)=\\ \nonumber &[(k_x-eBy)^2-\partial_y^2+k_z^2]^2+\alpha \mathbf{k}_F k_z^3 -\mathbf{k}_F ^4. 
\end{align}
This has the same eigenstates as a two dimensional electron gas in magnetic field, these eigenstates are well known\cite{Mahan2000}.The spectrum for states in the bulk are given by
\begin{equation}
E(B,n,k_z)=[\omega_c(n+\frac{1}{2})+k_z^2]^2+\alpha \mathbf{k}_F k_z^3 -\mathbf{k}_F ^4 
\end{equation}
where $\omega_c=e B$ is the cyclotron frequency. Since the vector potential $A_x(y)=B y$ is not periodic in $y$-drection, we will consider the system to
to be open along the $y$ direction with width $W$ 
and have periodic boundary conditions along the $x$ and $z$ direction. For this system the bulk extends for a 
range of $|k_x|<W/2 e B$ beyond which the bulk states merge into chiral edge states.
  Assuming that the system to be terminated in the $y$-drection by a potential $V(y)$, which varies smoothly 
on the scale of the magnetic length, the dispersion including both bulk and edge states is given by 
\begin{equation}\label{26}
E(B,n,k_x,k_z)=[\omega_c(n+\frac{1}{2})+k_z^2]^2+\alpha \mathbf{k}_F k_z^3 -\mathbf{k}_F ^4+V\left(\frac{k_x}{e B}\right). 
\end{equation}

The mean current carried by the system along the $z$-direction in steady state can be written as 
\begin{align}\label{30}
&\langle \mathbf{j}_z \rangle=-e\sum_{n,k_x}\int_{BZ}\frac{dk_z}{2\pi}\frac{\partial E(B,n,k_x,k_z)}{\partial k_z}f_n(k_x,k_z),
\end{align}
where $\frac{\partial E(B,n,k_x,k_z)}{\partial k_z}$ is the group velocity of the electrons along the $z$-direction and $f_n(k_x,k_z)$ 
is the occupation of the electronic states in the $n^{th}$ Landau level at wave-vector $k_x,k_z$.
For simplicity, we consider a system starting at a finite uniform magnetic field $B=B_1$. At such a finite magnetic field $B$, 
the Landau levels indexed by $n$ at any given momentum point $(k_x,k_z)$ are separated in energy and adiabatically increasing  
the magnetic field $B$ from $B=B_1$ to $B=B_2$ preserves the initially equilibrium occupation of the electronic levels which is given by 
\begin{align}\label{30a}
&f_n(k_x,k_z)=n_F(E(B_1,n,k_x,k_z)),
\end{align}
where $n_F(E)$ is the Fermi function at some temperature $T$.

It should be noted that as the magnetic field is raised the distribution no longer remains an equilibrium distribution.
In fact, the current can be shown to vanish in equilibrium in complete agreement with  Refs .\onlinecite{Vazifeh2013,Yamamoto2015}
since 
 \begin{align}\label{31}
&\expect{\mathbf{j}_z}=-e\sum_{n,k_x}\int_{BZ}\frac{dk_z}{2\pi}\frac{\partial \tilde{n_F}(E(B,n,k_x,k_z))}{\partial k_z}.
\end{align}
where $\tilde{n_F}(x)=\int_{-\infty}^x dx' n_F(x')$ is the integrated Fermi function. Noting that this function must approach a 
constant at the edge of the BZ where $n_F=0$,  the current density vanishes as   
$\expect{\mathbf{j}_z}=-e\sum_{n,k_x}[\tilde{n_F}(E\rightarrow \infty)-\tilde{n_F}(E\rightarrow \infty)]=0$. 

On the other hand, in the limit of a small bit finite change in the magnetic field, the current density aquires a 
finite expectation value that can be expanded to lowest order in $(B_2-B_1)$ as 
\begin{align}
&\langle \mathbf{j}_z \rangle=-e(B_2-B_1)\sum_{n,k_x}\int_{BZ}\frac{dk_z}{2\pi}\frac{\partial^2 E(B,n,k_x,k_z)}{\partial B \partial k_z}|_{B=B_1}\nonumber\\
&n_F(E(B_1,n,k_x,k_z)).
\end{align}
Assuming the zero temperature limit, the above integral can be restricted to be between $k_z=k_{z,1}$ and $k_z=k_{z,2}$, 
which are the unperturbed Fermi points defined by $E(B,n,k_x,k_z)=0$.  
With this simplification, the current density is written as 
\begin{align}\label{33}
&\expect{\mathbf{j}_z}=-\frac{e}{2\pi}\sum_{n,k_x} (B_2-B_1)\nonumber\\
&[\partial_BE(B,n,k_x,k_{z2})-\partial_B E(B,n,k_x,k_{z1})]|_{B=B_1}.
\end{align}
Substituting in $E$ from equation \eqref{26} gives
\begin{align}\label{34}
&\expect{\mathbf{j}_z}=-\frac{e^2}{\pi}\sum_{n,k_x} (n+\frac{1}{2})(B_2-B_1) (k_{z2}^2-k_{z1}^2)
\end{align}
Using $k_{z1}$,$k_{z2}$ to first order in $\alpha$ we obtain  
\begin{align}\label{34a}
&\expect{\mathbf{j}_z}=\frac{e^2}{\pi}\alpha k_F(B_2-B_1)\nonumber\\
&\sum_{n,k_x} \frac{(n+\frac{1}{2}) \Big[\big(k_F^4-V(\frac{k_x}{eB})\big)^{1/2}-eB_1(n+\frac{1}{2})\Big]^{3/2}}{\big(k_F^4-V(\frac{k_x}{eB})\big)^{1/2}}
\end{align}
which is nonzero in general even though the original Hamiltonian has no Berry curvature.

\section{Linear response in the clean systems}\label{III}
\subsection{Vanishing of low-frequency linear response for periodic boundary conditions}\label{IIIA}
In apparent contradiction to the previous section, the dynamical linear response of the current to a low frequency magnetic field has been shown to vanish.
To facilitate a direct comparison with our example, we review the argument in some detail. The key ingredient in this argument is to consider the response function 
in thermal equilibrium referred to as the equilibrium response, which is distinct from the DC limit of the dynamical response in general. The DC limit of the dynamical 
response is that real frequency response with the frequency being finite but small.

The response of the current operator  $\mathbf{j}(\mathbf{r})$ in thermal equilibrium to linear order in an external magnetic field  is given by 
\begin{align}\label{t1}
\langle \hat{j}_a(\mathbf{r})\rangle=\frac{Tr[\hat{j}_a(r) e^{-\beta (\hat{H}_0+\int d\mathbf{r}' \mathbf{j}(r').\mathbf{A}(r'))}]}{Tr[e^{-\beta (\hat{H}_0+\int d\mathbf{r}' \mathbf{j}(r').\mathbf{A}(r'))}]} + \langle \frac{\delta\hat{j}_a(\mathbf{r})}{\delta B}\rangle_0\
\end{align}
where the second term counts for the inrinsic change of the current operator $\mathbf{j}$ due to the application of the magnetic field.
 Here $\mathbf{A}(\mathbf{r})$ is the vector potential generated by the magnetic field and $\beta=1/k_B T$ is the inverse temperature. Defining 
\begin{align}
 &\hat{u}(\beta)=e^{-\beta (\hat{H}_0+\int d\mathbf{r}' \mathbf{j}(r').\mathbf{A}(r'))} e^{\beta \hat{H_0}}=\\ \nonumber 
 &1+\int dr' \int^\beta_0d\tau e^{-\tau \hat{H}_0}\mathbf{j}(r').\mathbf{A}(r')e^{\tau \hat{H}_0}+O(A^2)
\end{align}
and using it to expand equation \eqref{t1} first order gives 
\begin{align}
&\langle \hat{j}_a(\mathbf{r})\rangle=\langle \hat{j}_a(\mathbf{r})\rangle_0+\langle \frac{\delta\hat{j}_a(\mathbf{r})}{\delta B}\rangle_0\\ \newline \nonumber
&+\frac{Tr[\int dr' \int^\beta_0d\tau \hat{j}_a(r)e^{-\tau \hat{H}_0}\mathbf{j}(r').\mathbf{A}(r')e^{(\tau-\beta) \hat{H}_0}]}{Tr[e^{-\beta \hat{H}_0}]} \\ \newline \nonumber
&-\langle \hat{j}_a(\mathbf{r})\rangle_0 \frac{Tr[\int dr' \int^\beta_0d\tau e^{-\tau \hat{H}_0}\mathbf{j}(r').\mathbf{A}(r')e^{(\tau-\beta) \hat{H}_0}]}{Tr[e^{-\beta \hat{H}_0}]}\label{eq:j}
\end{align}
The first and the fourth terms evaluate to zero since $\langle \hat{j}_a(\mathbf{r})\rangle_0=0$. For a translational invariant system we can write 
\begin{align}
&\langle \hat{j}_a(\mathbf{r})\rangle=\frac{Tr[\Big(\int dr' \int^\beta_0d\tau \hat{j}_a(r)e^{-\tau \hat{H}_0}\mathbf{j}(r')_be^{(\tau-\beta) \hat{H}_0}\Big)]}{Tr[e^{-\beta \hat{H}_0}]}\mathbf{A}_b(r') \\ \newline \nonumber &+\langle \frac{\delta\hat{j}_a(\mathbf{r})}{\delta B}\rangle_0=\int d\mathbf{r'} \Pi_{ab}(\mathbf{r}-\mathbf{r'}) A_b(\mathbf{r'})+\langle \frac{\delta\hat{j}_a(\mathbf{r})}{\delta B}\rangle_0
\end{align}

Translational invariance suggests that the transformation to Fourier domain would simplify the results. However, it turns out that periodic boundary conditions restrict us from using  a 
strictly uniform magnetic field and we must consider a magnetic field with a finite but small wave-vector $\mathbf{q}$. At such a wave-vector we can readily choose the Fourier 
transform of the vector potential l $\mathbf{A}(r)$ to be $\mathbf{A}(\mathbf{q})=\frac{i}{q^2}\mathbf{B}\times\mathbf{q}$.
Using this, we can obtain the response of the lowest Fourier components of the current density as 
\begin{align}\label{tr1}
\langle \hat{j}_a(\mathbf{q})\rangle=&\frac{Tr[\Big(\int^\beta_0d\tau \hat{j}_a(\mathbf{q})e^{-\tau \hat{H}_0}\mathbf{j}(-\mathbf{q})_be^{(\tau-\beta) \hat{H}_0}\Big)]}{Tr[e^{-\beta \hat{H}_0}]}\mathbf{A}_b(\mathbf{q})\\ \newline \nonumber =&\sum_b \Pi_{ab}(\mathbf{q})\mathbf{A}_b(\mathbf{q}) +\langle \frac{\delta\hat{j}_a(\mathbf{\mathbf{q}})}{\delta B}\rangle_0
\end{align}
The second term vanishes in many of our examples and will be assumed to be zero for simplicity in the remainder of this section.
By relating the vector potential to the magnetic field,   we can rewrite equation \eqref{tr1} as 
\begin{equation} \label{tr2}
\langle \hat{j}_c(\mathbf{q})\rangle=\Big( \epsilon_{abc} \frac{1}{2iq} (\Pi_{ab}^R)^{ant}\Big) \mathbf{B}_c\equiv \sigma_{ch}\mathbf{B}_c.
\end{equation}
Expanding the current operator in terms of the creation operators  $\hat{c}^{\dagger}_{n,\mathbf{k}}$ for eigenstates $|m,\mathbf{k}\rangle$ of the Bloch Hamiltonian with eigenvalues $\varepsilon_{n,\mathbf{k}}$ 
as   
\begin{align}
\hat{j}_a(\mathbf{q})=\sum_{n,m,\mathbf{k}}\langle n,\mathbf{k}-\frac{\mathbf{q}}{2}|\hat{J}_a(\mathbf{k})|m,\mathbf{k}+\frac{\mathbf{q}}{2}\rangle \hat{c}^{\dagger}_{n,\mathbf{k}-\frac{\mathbf{q}}{2}} \hat{c}_{m,\mathbf{k}+\frac{\mathbf{q}}{2}}
\end{align}
where $\hat{J}_a(\mathbf{k})$ are the single particle current operators that are derived from the Bloch Hamiltonian. The response function 
 $\Pi_{ab}$ can be expanded as is standard with the derivation of the Kubo formula \cite{Mahan2000} to obtain a form 
\begin{align}\label{t2}
&\Pi_{ab}(\mathbf{q})=e^2\sum_{n,m,k}\frac{n_F(\varepsilon_{n,\mathbf{k}-\frac{\mathbf{q}}{2}})-n_F(\varepsilon_{m,\mathbf{k}+\frac{\mathbf{q}}{2}})}{\varepsilon_{n,\mathbf{k}-\frac{\mathbf{q}}{2}}-\varepsilon_{m,\mathbf{k}+\frac{\mathbf{q}}{2}}} \\ \nonumber
\times& \langle n,\mathbf{k}-\frac{\mathbf{q}}{2}|\hat{J}_a(\mathbf{k})|m,\mathbf{k}+\frac{\mathbf{q}}{2}\rangle \langle m,\mathbf{k}+\frac{\mathbf{q}}{2} |\hat{J}_b(\mathbf{k})|n,\mathbf{k}-\frac{\mathbf{q}}{2}\rangle.
\end{align}
This expression is identical to the result obtained for the dynamical linear response formalism in the limit $\omega\rightarrow 0$ or more precisely $\omega\ll q$.

Following the arguments of Refs. \onlinecite{Vazifeh2013,Yamamoto2015} it is easy to show that the result of equation \eqref{t1} and subsequently equation \eqref{t2} has to vanish as $\omega\rightarrow 0$ i.e. for 
as the magnetic field is varied slowly compared to $\mathbf{q}$. Therefore we conclude that 
\begin{align}\label{hl1}
\Pi_{ab}(\omega\ll q\rightarrow 0)=0
\end{align}

This result is in agreement with Ref \onlinecite{Chang2015} but in contrast to findings of Ref \onlinecite{Goswami2013}.

\subsection{Comparison with field theory results for Weyl semimetals}\label{IIIB}

One of the questions raised by the previous subsection is how to reconcile field theory predictions of a nonzero chiral magnetic response with our vanishing results. To investigate this we explicitly calculate equation \eqref{t2} for a generic two band model and use the result to calculate $\sigma_{ch}$ defined in equation \eqref{tr2} (details of this calculation are presented in the appendix).The final expression is given by 

\begin{align}\label{tr3}
&\sigma_{ch} =e\sum_{n=\pm}\int_{BZ} \frac{d\mathbf{k}}{(2\pi)^3} \nabla_{\mathbf{k}}.\mathbf{m}_n(\mathbf{k})f(\varepsilon_{n}(\mathbf{k}),t)   \\ \nonumber \newline
+&e^2 \sum_{n=\pm}\int_{BZ} \frac{d\mathbf{k}}{(2\pi)^3} (\frac{\mathbf{v}_{-,\mathbf{k}}+\mathbf{v}_{+,\mathbf{k}}}{2}).\mathbf{\Omega}_{n,\mathbf{k}}f(\varepsilon_{n}(\mathbf{k}),t)
\end{align}
where 
\begin{equation}\label{3}
\mathbf{m}_n(\mathbf{k})=-i\frac{e}{2}(\nabla_{\mathbf{k}}\langle n,\mathbf{k}|)\times[\hat{H}(\mathbf{k})-\varepsilon_n(\mathbf{k})](\nabla_{\mathbf{k}}|n,\mathbf{k}\rangle)
\end{equation}
 is the wave packet orbital magnetization. Our result for $\sigma_{ch}$ is in agreement with Ref.\onlinecite{Chang2015} but different from Refs.\onlinecite{Goswami2013}. Using periodicity of the lattice the second term in equation \eqref{tr3} can be partial integrated to look like the first term with the opposite sign therefore giving a vanishing $\sigma_{ch}$ as expected. However if we work within a low energy effective hamiltonian description of the problem, as is usually done in field theory calculations , a non vanishing result might have been achieved.To illustrate this point consider the simplest low energy effective hamiltonian of a Weyl semimetal, that is two linearly dispersing well fermions  (i.e. $H^{eff}=\pm v_F \sigma.\mathbf{k}$), in this case at each $\mathbf{k}$ is momentum space
 $\mathbf{v}_{+,\mathbf{k}}=-\mathbf{v}_{+,\mathbf{k}}=v_F \hat{k}$ and therefore the second term in equation \eqref{tr3} identically vanishes and we are left with 
 \begin{align}\label{tr4}
\sigma^{eff}_{ch} =e\sum_{n=\pm}\int_{BZ} \frac{d\mathbf{k}}{(2\pi)^3} \nabla_{\mathbf{k}}.\mathbf{m}_n(\mathbf{k})f(\varepsilon_{n}(\mathbf{k}),t)  
\end{align}

Partial integrating equation \eqref{tr4} in zero temperature gives
\begin{equation}\label{zx}
\sigma^{eff}_{ch} =e\int_{FS} \frac{d\mathbf{a}.\mathbf{m}_+(\mathbf{k})}{(2\pi)^3} 
\end{equation}
where $+$ here corresponds to the conduction band.
For a general two band Bloch Hamiltonian $H(\mathbf{k})=e(\mathbf{k})+\mathbf{r}(\mathbf{k})\cdot\mathbf{\sigma}$ (where 
$\sigma_{x,y,z}$ are the Pauli matrices) the energy eigenvalues are given by
\begin{equation}
\varepsilon_{\pm} =e(\mathbf{k})\pm |\mathbf{r}(\mathbf{k})|
\end{equation}
Substituting the eigenvalues and eigenvectors into Eq.~\ref{3}, the orbital magnetic moment is written as:
\begin{equation}
\mathbf{m}_{\pm}(\mathbf{k})=\pm e |\mathbf{r}(\mathbf{k})| \mathbf{\Omega} (\pm,\mathbf{k})\end{equation}
where $\mathbf{\Omega} (\pm,\mathbf{k})$ is the Berry curvature. Using this we can rewrite equation \eqref{zx} as 
\begin{align} 
\sigma^{eff}_{ch} =&e^2\int_{FS} |\mathbf{r}(\mathbf{k})| \frac{d\mathbf{a}\cdot\mathbf{\Omega} (+,\mathbf{k})}{(2\pi)^3} \label{fed}\\ \nonumber 
=&e^2\int_{FS} (\varepsilon_F-e(\mathbf{k})) \frac{d\mathbf{a}\cdot\mathbf{\Omega} (+,\mathbf{k})} {(2\pi)^3} \\ \nonumber 
=&-e^2\int_{FS} e(\mathbf{k}) \frac{d\mathbf{a}\cdot\mathbf{\Omega} (+,\mathbf{k})} {(2\pi)^3}
\end{align}
where we used the fact that the total Chern number of the entire Fermi surface is zero to get from the first line to the second line.
In case of a two node Weyl semimetal we have two  Fermi surfaces with $e_2-e_1=\delta k_0$ and opposite values of uniform Berry curvature  $\mathbf{\Omega} (+,\mathbf{k})$ therefore $\sigma^{eff}_{ch}=(\frac{e}{2\pi})^2 \delta k_0$ as expected from 
field theory \cite{Goswami2013a,Zyuzin2012}. This argument can be easily generalized to include an arbitrary number of Weyl nodes. Note that even though we recovered the quantum field theory result, it is not applicable to a 
real material since it doesn't include the second term in equation \eqref{tr3}. This term in a periodic system forces the chiral response $\sigma_{ch}=0$.

As has been pointed out, this situation can be partially circumvented by the chiral magnetic response at non zero frequencies where $\omega\gtrsim q$. 
While this limit can produce non-vanishing results even in lattice systems, it is difficult to disentangle the contribution of the electric field generated by the 
time-dependence of the magnetic field in this limit. The finite frequency generalization of the linear response calculation is given by \cite{Mahan2000} 
\begin{align}\label{w1}
&\Pi_{ab}(\mathbf{q},\omega)=e^2\sum_{n,m,k}\frac{f(\varepsilon_{n,\mathbf{k}-\frac{\mathbf{q}}{2}})-f(\varepsilon_{m,\mathbf{k}+\frac{\mathbf{q}}{2}})}{\omega+\varepsilon_{n,\mathbf{k}-\frac{\mathbf{q}}{2}}-\varepsilon_{m,\mathbf{k}+\frac{\mathbf{q}}{2}}} \\ \nonumber
\times& \langle n,\mathbf{k}-\frac{\mathbf{q}}{2}|\hat{J}_a(\mathbf{k})|m,\mathbf{k}+\frac{\mathbf{q}}{2}\rangle \langle m,\mathbf{k}+\frac{\mathbf{q}}{2} |\hat{J}_b(\mathbf{k})|n,\mathbf{k}-\frac{\mathbf{q}}{2}\rangle
\end{align}
to investigate behavior of $\sigma_{ch}$ as a function of frequency, we will numerically calculate  equation \eqref{w1}  for a simple model two band model of Weyl semimetal with two Weyl nodes at zero temperature.
In this calculation, the vanishing of the $\omega\ll q\rightarrow 0$ response comes to our aid and we can use this fact to argue that the inter-band terms must cancel with the $\omega\rightarrow 0$ limit of the 
intra-band (i.e. Fermi surface) terms. Therefore, the finite frequency response only Fermi surface properties contribute to equation \eqref{w1} and therefore no further knowledge of the microscopic details of the Hamiltonian are necessary. Focusing on the intraband contribution to Eq.~\ref{w1} we obtain 
\begin{align}\label{y1}
&\Pi_{ab}^{intra}(\mathbf{q},\omega)=e^2\sum_{n,k}\frac{\theta(\varepsilon_{n,\mathbf{k}-\frac{\mathbf{q}}{2}})-\theta(\varepsilon_{n,\mathbf{k}+\frac{\mathbf{q}}{2}})}{\omega+\varepsilon_{n,\mathbf{k}-\frac{\mathbf{q}}{2}}-\varepsilon_{n,\mathbf{k}+\frac{\mathbf{q}}{2}}} \\ \nonumber
\times& \langle n,\mathbf{k}-\frac{\mathbf{q}}{2}|\hat{J}_a(\mathbf{k})|n,\mathbf{k}+\frac{\mathbf{q}}{2}\rangle \langle n,\mathbf{k}+\frac{\mathbf{q}}{2} |\hat{J}_b(\mathbf{k})|n,\mathbf{k}-\frac{\mathbf{q}}{2}\rangle.
\end{align}

\begin{figure}
\centering
\includegraphics[width=\columnwidth,height=\textheight,keepaspectratio]{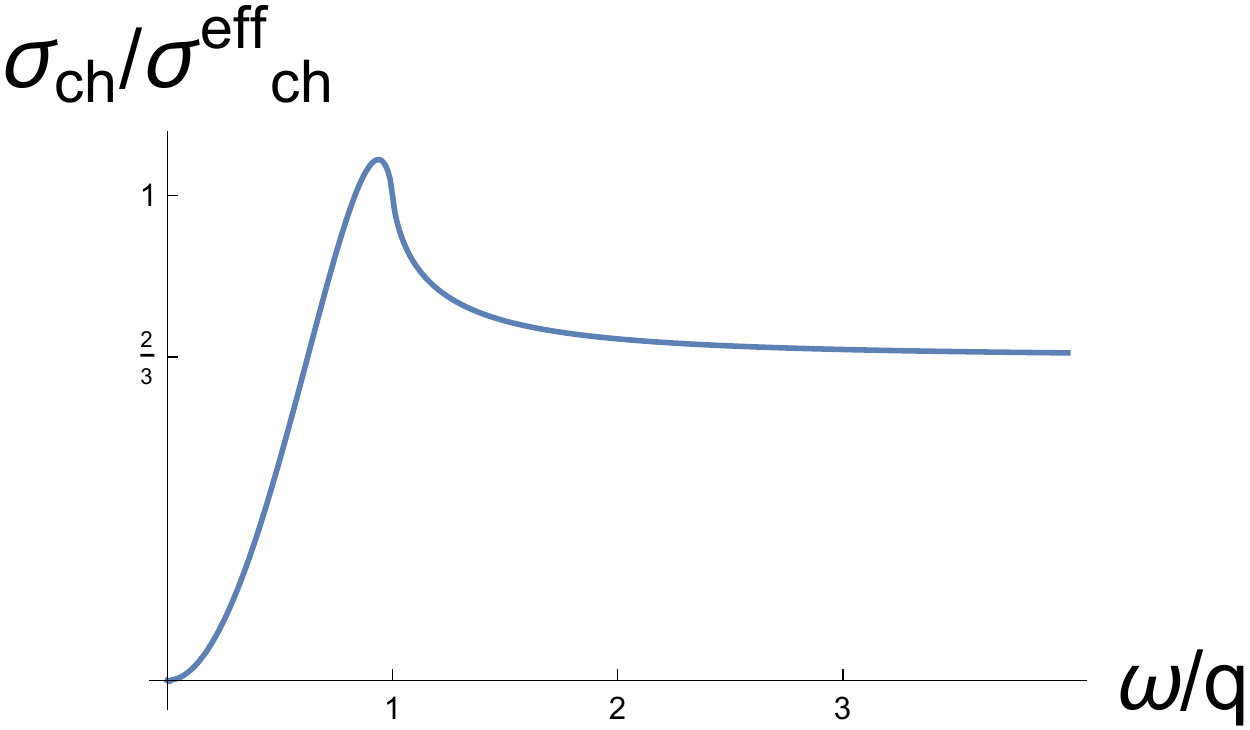} \caption{Frequency dependence of the Chiral Magnetic response $\sigma_{ch}(q,\omega)$  in a bulk Weyl semimetal in a two band two nodes model. It vanishes in the DC (i.e. $\omega\rightarrow 0$) limit as expected from the equilibrium theory. 
We chose the parameter $q=0.0001$ \label{fig:w2.pdf}}
\end{figure}

The results of this calculation are plotted in figure \ref{fig:w2.pdf}. It starts from zero at $\omega \ll q$ as expected from equilibrium theory, then peaks at some frequency an then approaches $\sigma_{ch}=\frac{2}{3} \sigma_{ch}^{eff}$ at $\omega \gg q$ note however that since $\mathbf{E}\propto \omega \mathbf{A}$ and $\mathbf{B}\propto q \mathbf{A}$ in this limit $E\gg B$ and therefore nonzero $\sigma_{ch}$ in this limit is more of an electric field effect rather than magnetic field one. It is worth mentioning that in the limiting case of $\omega=q\ll 1$ we get $\sigma_{ch}=\sigma_{ch}^{eff}$ we believe this feature is coincidental, since this limit does not correspond to $B\neq 0$ and $E=0$ as required in the DC chiral magnetic effect.

Also note that the topology of a Weyl semimetal is not necessary to obtain a nonzero $\sigma_{ch}$ \cite{Chang2015a,Ma2015,Moore}. One way to see this is to look at the limit $\omega\gg q$ direct calculation of equation \eqref{w1} for an isotropic model in this limit gives
\begin{align}
\sigma_{ch}=\frac{2}{3}\sigma_{ch}^{eff}=\frac{2}{3} e\sum_{n=\pm}\int_{BZ} \frac{d\mathbf{k}}{(2\pi)^3} \nabla_{\mathbf{k}}.\mathbf{m}_n(\mathbf{k})f(\varepsilon_{n}(\mathbf{k}),t)  
\end{align}
all the steps from equation \eqref{tr4} to equation \eqref{fed} goes through here as well. Interestingly $e(\mathbf{k})$ and $\mathbf{\Omega} (\pm,\mathbf{k})$ in Eq.\eqref{fed} are independent of each other since the Berry curvature only depends on eigenstates not eigenvalues. Therefore as long as Berry curvature is not zero every where on Fermi surface we can choose $e(\mathbf{k})$ arbitrarily such that $\sigma_{ch}$ is non zero. 
Therefore, similar to the magneto-electric effect \cite{Essin2009}, topology, which is defined by Fermi surface components 
with non-vanishing Berry flux ~\cite{Haldane2014} is not necessary to  a get nonzero $\sigma_{ch}$. Similar finite frequency CME resulting from non-topological Berry curvature has been previously reported \cite{Chang2015a,Ma2015,Moore}.

 \begin{figure*}[t]
\centering
\includegraphics[width=\textwidth,height=\textheight,keepaspectratio]{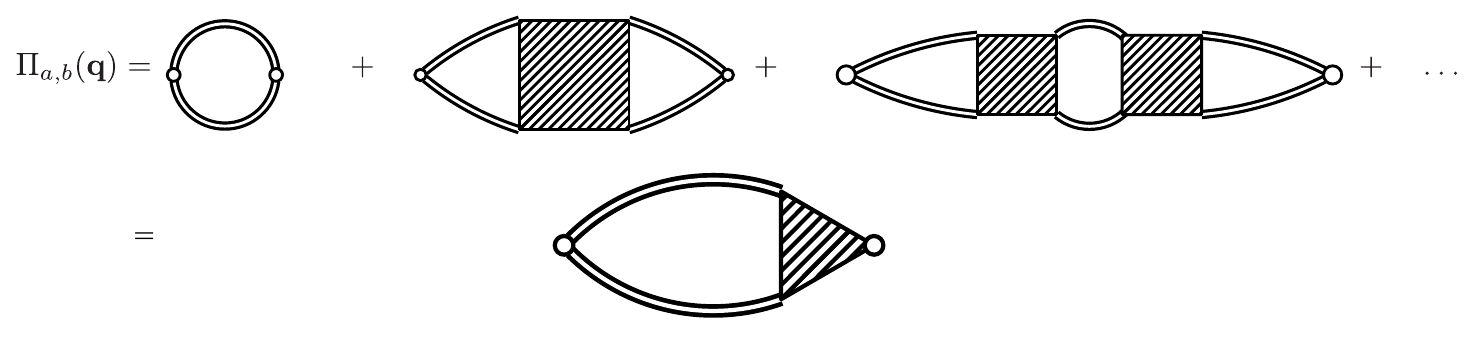} \caption{Feynman diagrams contributing to the correlator $\Pi_{a,b}$. Double lines correspond to $G(\mathbf{\widetilde{k}})$ (i.e. dressed propagator) and the shaded boxes correspond to two particle irreducible diagrams \cite{Mahan2000}\label{fig:merged1.pdf}}
\end{figure*}

\subsection{ $B$-field response under open boundary conditions}\label{IIIC}

The second issue raised by the vanishing DC limit of the dynamical response, which has not been resolved earlier, is the apparent contradiction 
 between equation \eqref{hl1} and the example presented in Sec.~\ref{II}. As we will show, the crux of this discrepancy lies in the fundamental 
difference in the description of the magnetic field for systems with open and periodic boundary conditions. 

Unlike in the case of periodic boundary conditions, where a magnetic field must be applied with a finite wave-vector $\mathbf{q}$, open 
systems can be subject to a strictly uniform magnetic field as in experiments. 
A uniform magnetic field $\mathbf{B}$ in an open system can be represented  in the circular gauge by a vector potential $\mathbf{A}$ that  is given by $\bm{A}=\frac{1}{2}\mathbf{B}\times \mathbf{r}$.
In this case, the magnetic field perturbation  $\delta B$ affects the Hamiltonian as $H\rightarrow H-\mathbf{M}_z \delta B$, 
where $\hat{M}_b=\frac{1}{2}\int d\mathbf{r} (\mathbf{J}(\mathbf{r})\times \mathbf{r})_b$ is the magnetic moment operator.
Using this in Eq.~\ref{eq:j} from Sec.~\ref{IIIA} and noting that $\int d\mathbf{r'}\mathbf{J(r')}\cdot\mathbf{A(r')}$ transforms 
to $M_z B$ in the present notation, the response of the equilibrium current density to magnetic field is 
\begin{align}
&\delta  \expect{j}=\int d\tau \frac{Tr[j e^{-(\beta-\tau)H_0}M_z e^{-\tau H_0}]}{Tr[e^{-\beta H_0}]},
\end{align}
where $j\equiv j_z$ is the current operator in the z direction.Expanding in the quasiparticle operator  eigenbasis $H_0=\sum_p \epsilon_p c_p^\dagger c_p$ 
and the other operators as $j=\sum_{p,q}J_{p,q}c_p^\dagger c_q$ and $M=\sum_{r,s}m_{r,s}c_r^\dagger c_s$ the current response matrix element becomes 
 \begin{align}
&\delta  \expect{j}=\sum_{p,q,r,s}J_{pq}m_{rs}\int d\tau\frac{Tr[ c_p^\dagger c_q e^{-(\beta-\tau)H_0} c_r^\dagger c_s e^{-\tau H_0}]}{Tr[e^{-\beta H_0}]}.
\end{align}
Noting that $e^{\tau H_0}c^\dagger_r e^{-\tau H_0}=c^\dagger_r e^{-\epsilon_r\tau}$ 
 \begin{align}
&\delta  \expect{j}=\sum_{p,q,r,s}J_{pq}m_{rs}\frac{Tr[ c_p^\dagger c_q e^{-\beta H_0} c_r^\dagger c_s ]}{Tr[e^{-\beta H_0}]}\int_0^\beta d\tau e^{-\tau (\epsilon_r-\epsilon_s)}.
\end{align}
Separating the  $r\neq s$ and $r=s$ contribution to the current response $\delta  \expect{j}=\delta  \expect{j}_{r\neq s}+\delta  \expect{j}_{r=s}$, the $r\neq s$ contribution is written as 
\begin{align}\label{interband}
&\delta\expect{j}_{r\neq s}=e^2\sum_{r\neq s}\frac{f(\varepsilon_{r})-f(\varepsilon_{s})}{\varepsilon_{r}-\varepsilon_{s}} \times {\langle r}|\hat{J}_z|s\rangle \langle s |\hat{M}_z|r\rangle.
\end{align}

This term is equivalent to the DC limit of the finite frequency linear response. To compare this result to the dynamical linear response in Eq.~\ref{t2} we notice that $k_{x,y}$ are no longer good quantum numbers
in the open boundary condition case and we can simply 
replace $\textbf{k}\rightarrow k_z$ in the derivation in Sec.~\ref{IIIA}. Therefore, the open system limit is obtained
from Eq.~\ref{t2} by dropping the $\mathbf{k,q}$ labels and is written as 
\begin{align}\label{t2z}
&\Pi_{z}=e^2\sum_{n,m}\frac{f(\varepsilon_{n})-f(\varepsilon_{m})}{\varepsilon_{n}-\varepsilon_{m}-\omega} \times {\langle n}|\hat{J}_z|m\rangle \langle m |\hat{M}_z|n\rangle.
\end{align}
Now note that at any finite $\omega>0$, the $m=n$ contribution to the above sum vanishes so that the DC (i.e. $\omega\rightarrow 0$) limit of this expression 
is identical to  that in Eq.~\ref{interband}.

In addition to the DC limit of the dynamic response, there is also a contribution to the equilibrium current response $\delta\expect{j}$, which is written as 
 \begin{align}
&\delta  \expect{j}_{r=s}=\beta\sum_{p,r}J_{pp}m_{rr} [f(\epsilon_r)f(\epsilon_p)(1-\delta_{pr})+\delta_{rp}f(\epsilon_r)],\label{eq:rs}
\end{align}
where we note that the $p=r$  and $p\neq r$ cases lead to different terms in the above expression and $f(\epsilon_p)=\expect{c_p^\dagger c_p}$ are Fermi functions.
In contrast, the analogue of the $r=s$ term does not contribute to the finite frequency response function. 
Finally, we note that the explicit response of the current $\expect{\frac{\delta j}{\delta B}}$ is identical in both real and imaginary frequency cases. We have ignored this contribution 
for simplicity.

The $r=s$ term in Eq.~\ref{eq:rs} can lead to a substantial difference between the vanishing equilibrium response and the DC limit of the dynamical response.
 As a result, while the equilibrium linear 
response is required to vanish based on the argument in Sec.~\ref{IIIA}, the DC limit (i.e. $\omega\rightarrow 0$)  of the dynamic response is not necessarily 
vanishing as suggested by Sec.~\ref{II}. This is consistent with the non-vanishing CME obtained for certain open systems \cite{Baireuther2015}.

\section{Chiral magnetic response in weakly disordered systems}\label{IV}

The necessity of a finite but small wave-vector $\mathbf{q}$ for the magnetic field used in the linear response derivation in Sec.~\ref{IIIA} leads to 
some subtle difficulties in the order of limits. 
 This is because that the vector potential scales as $A\approx \frac{B}{q}$ and therefore it diverges as $\mathbf{q}\rightarrow 0$, so that as $q\rightarrow 0$, the range of $B$ over which 
the perturbation theory is valid shrinks to zero. This difficulty can be avoided by introducing another length scale into the problem so that the response function becomes independent of 
$q$ at small enough $q$. One way to do this is to introduce the length $\frac{1}{\tau}$ given by the inverse of the scattering rate, in this case the wave vector $q$ just needs to be much smaller than the mean free path $q\ll \frac{1}{\tau}$ rather than going to zero $q \rightarrow 0$. To address this problem we'll consider the problem of static CME in a disordered metal in the last section.There we'll show that equation \eqref{hl1} remains valid in presence of weak disorder.

As mentioned in the introduction of disorder introduces a length scale to the the system $\frac{1}{q}$, that 
can help make the perturbation theory valid when the magnetic field is turned on. 
We introduce disorder into a lattice realization of a Weyl semi-metal through a potential term in the 
Hamiltonian, which is written as:
\begin{align}\label{111}
V=\sum_{{\bf r},a}u_a({\bf  r})c_{a}({\bf r})^\dagger c_{a}({\bf r})
\end{align}
where ${\bf r}$ labels unit cells and $a$ labels atoms inside the unit cell. 
For our calculations, we use a Gaussian white-noise disorder model for the functions $u_a(\bf r)$ with a 
correlation function 
$\expect{u_a({\bf r})u_b({\bf r}')}=\nu_D\delta_{a,b}\delta_{\bf r-r'},$ where $\nu$ characterizes the 
strength of the disorder. 
The potential perturbation $V$ in  Fourier space is written as  
\begin{align}
V=\sum_{{\bf k,q},a}u_a({\bf  q})c_{a}({\bf k+q})^\dagger c_{a}({\bf k})
\end{align}
where $\expect{u_a({\bf q})u_b^*({\bf q}')}=\nu_D\delta_{a,b}\delta_{\bf q-q'}$.
Starting with this perturbation, the disordered averaged Green function can be calculated within the 
Born approximation \cite{Mahan2000} as $G(\mathbf{\widetilde{k}})^{-1}=\omega-\hat{H}(\mathbf{k})-\Sigma(\omega)$ 
where 
\begin{align}
\Sigma_{ab}({\omega})=\nu_D\delta_{ab}\int d{\bf q}d\omega G_{aa}^{(0)}({\bf q},\omega)
\end{align}
is the electron self-energy within the Born approximation
and $G^{(0)}$ is the bare time-ordered Green  function (i.e. $G^{(0)}({\bf q},\omega)=[\omega+ i \sign(\omega)\eta-H({\bf k})]^{-1}$). Note that for compactness we have introduced the notation $\widetilde{k}:=(\mathbf{k},\omega)$.

To calculate the disorder averaged response $\sigma_{ch}$, we use the Kubo formula as in the clean case modified to 
include weak disorder. Following the standard diagrammatic theory for disorder~\cite{Mahan2000}, we do this by  calculating the Feynman diagrams shown in Fig [\ref{fig:merged1.pdf}]. In these diagrams, the double lines correspond to disorder -averaged Green functions $G(\mathbf{\widetilde{k}})$ and the shaded boxed correspond to disorder scattering by the fluctuations in the potential $V$. 
We can sum all of the contributing diagrams into a renormalized current vertex, $\Gamma_a$ (shown in the second line Fig [\ref{fig:merged1.pdf}]), so that the response function is written as:
\begin{align}\label{15}
&\Pi_{a,b}(\mathbf{q},\omega)=e^2Tr\big[\int_{-\infty}^{\infty}\frac{d\omega}{2\pi}\int_{BZ}\frac{d^3k}{(2\pi)^3}\hat{J}_a(\mathbf{k})\\ \nonumber
\times &G(\widetilde{\mathbf{k}}+\frac{\widetilde{\mathbf{q}}}{2})\mathbf{\Gamma}_b(\widetilde{\mathbf{k}}+\frac{\widetilde{\mathbf{q}}}{2},\widetilde{\mathbf{k}}-\frac{\widetilde{\mathbf{q}}}{2})G(\widetilde{\mathbf{k}}-\frac{\widetilde{\mathbf{q}}}{2})\big]
\end{align}
where 
\begin{align}
&\mathbf{\Gamma}_b(\widetilde{\mathbf{k}}+\frac{\widetilde{\mathbf{q}}}{2},\widetilde{\mathbf{k}}-\frac{\widetilde{\mathbf{q}}}{2})=\hat{J}_b(\mathbf{k})+\nu_D\int \frac{d\mathbf{\widetilde{q}}' }{2\pi}
\\ \nonumber \times
&G(\widetilde{\mathbf{k}}+\frac{\widetilde{\mathbf{q}}}{2}+\widetilde{q}') \mathbf{\Gamma}_b(\widetilde{\mathbf{k}}+\frac{\widetilde{\mathbf{q}}}{2}+\widetilde{q}',\widetilde{\mathbf{k}}-\frac{\widetilde{\mathbf{q}}}{2}-\widetilde{q}')G(\widetilde{\mathbf{k}}-\frac{\widetilde{\mathbf{q}}}{2}-\widetilde{q}')
\end{align}

We note that the validity of this approach  requires being in the diffusive limit (i.e. mean-free path $\gg$ Fermi wave-length). 
This is different from the Weyl semi-metal regime with vanishing density of states that is being debated  for chemical potential 
near the Weyl node \cite{Pixley2015c,Roy2014,Sbierski2014,Ominato2014,Pixley2015c,Bera2015}. We will avoid this  regime by choosing a finite chemical potential with a Fermi 
energy much greater than the disorder scattering rate.

In principle, once $\Pi$ is calculated using Eq.~\ref{15}, one can substitute it back into Eq.~\ref{tr2} to calculate the 
chiral magnetic response $\sigma_{ch}$. We now argue that this necessarily vanishes for a disordered system. To do this,
 note that the Ward identity \cite{Schrieffer1983} gives: 
\begin{eqnarray}
 \omega \Gamma_0(\widetilde{\mathbf{k}}+\frac{\widetilde{\mathbf{q}}}{2},\widetilde{\mathbf{k}}-\frac{\widetilde{\mathbf{q}}}{2})- \mathbf{q}.\mathbf{\Gamma}(\widetilde{\mathbf{k}}+\frac{\widetilde{\mathbf{q}}}{2},\widetilde{\mathbf{k}}-\frac{\widetilde{\mathbf{q}}}{2})\\ \nonumber=
-G^{-1}(\widetilde{\mathbf{k}}+\frac{\widetilde{\mathbf{q}}}{2})+G^{-1}(\widetilde{\mathbf{k}}-\frac{\widetilde{\mathbf{q}}}{2}),
\end{eqnarray}
which in turn guarantees that in the limit that we are interested in (i.e. $\frac{\omega}{q}\rightarrow 0$) :
\begin{equation}\label{a5a}
\mathbf{\Gamma}_a(\widetilde{\mathbf{k}}+\frac{\widetilde{\mathbf{q}}}{2},\widetilde{\mathbf{k}}-\frac{\widetilde{\mathbf{q}}}{2})= \partial_{k_a} G^{-1}(\widetilde{\mathbf{k}})=\partial_{k_a} \hat{H}(\mathbf{k})=\hat{J}_a(\mathbf{k}).
\end{equation}
This implies that there are no vertex corrections and we need to consider only the bubble diagram (the first diagram in the first line of Fig.~\ref{fig:merged1.pdf}). With this approximation, $\Pi$ in Eq.~\ref{15} when expanded to first order in $q$ becomes:
\begin{align}\label{18}
\Pi_{a,b}(\mathbf{q},\omega)=e^2&\frac{q}{6}\sum_{a,b,c}\varepsilon_{a,b,c}Tr\big[\int_{-\infty}^{\infty}\frac{d\omega}{2\pi}\int_{BZ}\frac{d^3k}{(2\pi)^3}  \\ \nonumber
\times &\partial_{k_a} G^{-1}(\widetilde{\mathbf{k}})\partial_{k_c}G(\widetilde{\mathbf{k}})\partial_{k_b} G^{-1}(\widetilde{\mathbf{k}})G(\widetilde{\mathbf{k}})\big]
\end{align}

This has the form of a Hopf topological invariant \cite{Fradkin2013} and vanishes since the Green function $G$ has no real frequency poles (shown in the appendix). Therefore, using Eq.~\ref{tr2}, we conclude that $\sigma_{ch}=0$ universally for all disordered physical systems. This is consistent with the equilibrium results showing that the current must vanish in a magnetic field in a 
lattice system~\cite{Vazifeh2013,Yamamoto2015}.

On the other hand, for frequencies much larger than the scattering rate $\omega>>\frac{1}{\tau}$, we expect disorder not to play a role and therefore finite frequency chiral magnetic response $\sigma_{ch}$ should return to the clean limit value 
in such a range. In this case, we have to be careful to ensure that the wave-vector $\mathbf{q}$ is chosen to obey the 
limit $\omega/q\rightarrow 0$.

\begin{figure}
\centering
\includegraphics[width=\columnwidth,height=\textheight,keepaspectratio]{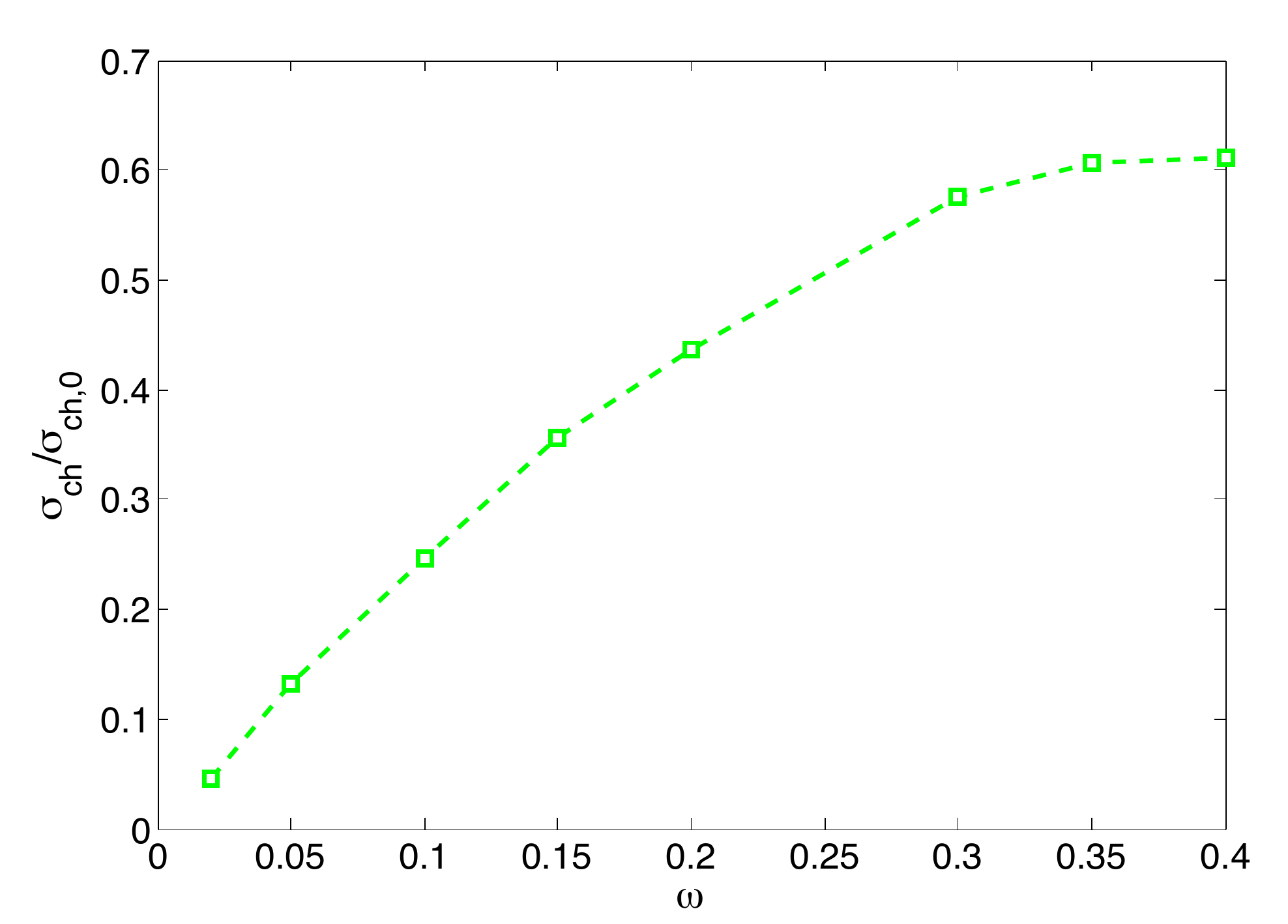} \caption{Frequency dependence of the Chiral Magnetic response $\sigma_{ch}(q,\omega)$  in a bulk Weyl semimetal. While $\sigma_{ch}$ reaches a value close to the 
clean limit of   $\approx 0.6\sigma_{ch,0}=0.6 t (e/2\pi)^2$ for frequencies exceeding the disorder scattering rate $\frac{1}{\tau}=0.05\propto \nu_D$, 
it vanishes in the DC (i.e. $\omega\rightarrow 0$) limit as expected from the equilibrium theory. 
For the calculation, we chose the parameter $t=0.15$ and the wave-vector $q=0.4$\label{fig:CMEnum.pdf}}
\end{figure}

To understand how $\sigma_{ch}$ crosses over from the vanishing DC  value to the clean-limit value, we numerically calculate $\Pi_{y,z}(q \hat{x},\omega)$ for the model Hamiltonian of Weyl semimetal used in Ref.\onlinecite{Goswami2013} :
\begin{align}\label{cc}
&\hat{H}=\sum_{\mathbf{k}}\Psi^{\dagger}_{\mathbf{k}}[N_{0,\mathbf{k}}\sigma_0+\mathbf{N}_{\mathbf{k}}.\mathbf{\sigma}]\Psi_{\mathbf{k}} \\ \nonumber
&N_{0,\mathbf{k}}=8 t\prod_i \cos (k_i) \\ \nonumber
&N_{j,\mathbf{k}}=\sin (k_j)
\end{align}
This model hosts four right handed Weyl Fermions located at high symmetry points $(0,0,0), (\pi,\pi,0) ,(\pi,0,\pi), (0,\pi,\pi)$ and four left handed ones at $(\pi,\pi,\pi),(0,0,\pi),(0,\pi,0),(\pi,0,0)$ . We use the disorder realization as in equation \eqref{111} and 
use Eq.~\ref{15} to calculate  $\Pi_{y,z}(q \hat{x},\omega)$ so that we can calculate $\sigma_{ch}$ using Eq.~\eqref{a5a} 
(without taking the limits $\omega,q\rightarrow 0$).
To reduce the numerical complexity we assume that $\omega/q$ is still small enough so that we can use Eq.~\eqref{a5a}.
Within this approximation, we then replace the self-energy by a uniform scattering rate $\Sigma(\omega)\approx i\tau^{-1}=i0.05$.
The resulting $\sigma_{ch}(\omega)$ from our calculation, which is plotted in Fig.~\ref{fig:CMEnum.pdf}, shows 
that $\sigma_{ch}$ vanishes in the DC limit and approaches $\approx 0.6\sigma_{ch,0}=0.6 t (e/2\pi)^2$, which is consistent with the 
clean limit for the chosen $q=0.4$. 

\section{Conclusion}
In summary, we have shown that a CME-like response i.e. one where a current flows in response to a magnetic field is 
in principle possible with or without Weyl nodes. This appears to contradict previous claims of the vanishing of the low frequency CME. 
We point out that the derivation of the vanishing CME is a consequence of periodic boundary conditions of the system. A more realistic 
system with open boundary conditions would not be subject to the same constraints and can have a non-vanishing CME.
We also studied the finite frequency CME with periodic boundary conditions and consistent with recent work, we found it to be 
non-vanishing in general when there was a non-vanishing Berry curvature on the Fermi surface. This does not necessitate having a 
topological Berry flux as in the case of a Weyl node.
Finally, we study how the perturbation theory in magnetic field might be more stable in the presence of disorder.
Using the standard diagrammatic treatment of 
disorder within the Born approximation, we have found that in a realistic disordered system, the chiral magnetic 
response is really a dynamical phenomena and vanishes in the DC limit. For frequencies in excess of the scattering rate, 
the clean limit predictions are recovered. Numerical evaluation of the associated integrals for a specific lattice model 
show how the cross-over occurs as the frequency is increased above the scattering rate.

This work was supported by the JQI-NSF-PFC and start-up funds from the University of Maryland. We further thank Dima Pesin, Joel Moore, 
Ivo Souza for pointing out an error in the original version of this work.

\bibliographystyle{h-physrev}
\bibliography{library}

\appendix*
\begin{widetext}
\section{Appendices}
\subsection{\label{sec:level1}Details of the clean linear response calculation}
Here we explicitly show how to get from equations \eqref{tr2} and \eqref{t2} to equation \eqref{tr3}. In the limit we are interested in (i.e. $\lim_{q\rightarrow 0} \lim_{\omega\rightarrow 0}$) we can re write equation \eqref{t2} as :
\begin{align} 
\Pi_{ab}^R(\mathbf{q},\omega)=e^2&\sum_{n,m\neq n,k}\frac{f(\varepsilon_{n,\mathbf{k}-\frac{\mathbf{q}}{2}})-f(\varepsilon_{m,\mathbf{k}+\frac{\mathbf{q}}{2}})}{\varepsilon_{n,\mathbf{k}-\frac{\mathbf{q}}{2}}-\varepsilon_{m,\mathbf{k}+\frac{\mathbf{q}}{2}}}  \langle n,\mathbf{k}-\frac{\mathbf{q}}{2}|\hat{J}_a(\mathbf{k})|m,\mathbf{k}+\frac{\mathbf{q}}{2}\rangle \langle m,\mathbf{k}+\frac{\mathbf{q}}{2} |\hat{J}_b(\mathbf{k})|n,\mathbf{k}-\frac{\mathbf{q}}{2}\rangle \\ \nonumber 
+e^2&\sum_{n,k}f'(\varepsilon_{n,\mathbf{k}})\langle n,\mathbf{k}-\frac{\mathbf{q}}{2}|\hat{J}_a(\mathbf{k})|m,\mathbf{k}+\frac{\mathbf{q}}{2}\rangle \langle m,\mathbf{k}+\frac{\mathbf{q}}{2} |\hat{J}_b(\mathbf{k})|n,\mathbf{k}-\frac{\mathbf{q}}{2}\rangle
\end{align}

We now expand to first order in $\mathbf{q}$ and keep only the anti-symmetric part, for simplicity we divide the expression to four terms each corresponding to the expansion of :\\
1.the numerator of the first term.$(\Pi_1$)\\
2.the denominator of the first term.$(\Pi_2)$\\
3.the matrix element in the first term.$(\Pi_3)$\\
4.the matrix element in the second term.$(\Pi_4)$\\
Now we calculate each one as  follows ($\mathbf{v}_{n,\mathbf{k}}=\nabla_{\mathbf{k}}\varepsilon_{n,\mathbf{k}}$ everywhere below):

\begin{align}
(\Pi_1)^{ant}=-e^2\sum_{n,m\neq n,k}f'(\varepsilon_{n,\mathbf{k}})(\mathbf{v}_{n,\mathbf{k}}.\frac{\mathbf{q}}{2})\frac{ \langle n,\mathbf{k}|\hat{J}_a(\mathbf{k})|m,\mathbf{k}\rangle \langle m,\mathbf{k} |\hat{J}_b(\mathbf{k})|n,\mathbf{k}\rangle}{\varepsilon_{n,\mathbf{k}}-\varepsilon_{m,\mathbf{k}}} = i e^2 \sum_{n,m\neq n,k}f'(\varepsilon_{n,\mathbf{k}})(\mathbf{v}_{n,\mathbf{k}}.\mathbf{q})  (\mathbf{m}_n(\mathbf{k}))_c
\end{align}

after shifting $\mathbf{k}$ to $\mathbf{k}-\frac{\mathbf{q}}{2}$, for $\Pi_2$ we have:

\begin{align}
(\Pi_2)^{ant}=ie^2\sum_{n,m\neq n,k}f(\varepsilon_{n,\mathbf{k}})\frac{(\mathbf{v}_{n,\mathbf{k}}+\mathbf{v}_{m,\mathbf{k}}).\mathbf{q}}{2}\frac{ \langle n,\mathbf{k}|\hat{J}_a(\mathbf{k})|m,\mathbf{k}\rangle \langle m,\mathbf{k} |\hat{J}_b(\mathbf{k})|n,\mathbf{k}\rangle}{(\varepsilon_{n,\mathbf{k}}-\varepsilon_{m,\mathbf{k}})^2} =ie^2 \sum_{n,k}f(\varepsilon_{n,\mathbf{k}})\frac{(\mathbf{v}_{n,\mathbf{k}}+\mathbf{v}_{m,\mathbf{k}}).\mathbf{q}}{2}  (\mathbf{\Omega}_n(\mathbf{k}))_c
\end{align}

. Calculation of $\Pi_3$ is rather complicated and in order to get a closed form we assume that the model has only two bands, after a rather lengthy calculation we get : 
\begin{align}
(\Pi_3)^{ant}=ie^2 \sum_{n,k}f(\varepsilon_{n,\mathbf{k}})|\mathbf{q}|\Big[(\mathbf{v}_{+,\mathbf{k}}+\mathbf{v}_{-,\mathbf{k}})_a(\mathbf{\Omega}_n(\mathbf{k}))_a +(\mathbf{v}_{+,\mathbf{k}}+\mathbf{v}_{-,\mathbf{k}})_b(\mathbf{\Omega}_n(\mathbf{k}))_b \Big]
\end{align}
Finally for $\Pi_4$ we have ( after expanding the matrix elements and simplifying ):

\begin{align}
(\Pi_4)^{ant}=i e^2 \sum_{n,k}f'(\varepsilon_{n,\mathbf{k}}) |\mathbf{q}|\Big[(\mathbf{v}_{n,\mathbf{k}})_a(\mathbf{m}_n(\mathbf{k}))_a +(\mathbf{v}_{n,\mathbf{k}})_b(\mathbf{m}_n(\mathbf{k}))_b \Big]
\end{align}
After putting every thing together and changing the sum into an integral we get :

\begin{align}
\sigma_{ch} =\lim_{q \rightarrow 0} \epsilon_{abc} \frac{1}{2iq} \lim_{\omega \rightarrow 0} (\Pi_{ab}^R(\mathbf{q},\omega))^{ant}=&-e\sum_n\int_{BZ} \frac{d\mathbf{k}}{(2\pi)^3} (\mathbf{v}_{n,\mathbf{k}}.\mathbf{m}_n(\mathbf{k}))f'(\varepsilon_{n}(\mathbf{k}),t) \\ \nonumber \newline&+ e^2 \sum_{n=\pm}\int_{BZ} \frac{d\mathbf{k}}{(2\pi)^3} (\frac{\mathbf{v}_{-,\mathbf{k}}+\mathbf{v}_{+,\mathbf{k}}}{2}).\mathbf{\Omega}_{n,\mathbf{k}}f(\varepsilon_{n}(\mathbf{k}),t)
\end{align}
Which after partial integrating becomes Equation\eqref{tr3}, Note that because of the assumptions made in  calculating $\Pi_2$ and $\Pi_3$ this result is only valid for a two band model.

\subsection{\label{sec:level2}Universal vanishing of the Hopf term}
To prove that Equation \eqref{18} is really a topological invariant we consider the effect of changing the Hamiltonian from $\hat{H}(\mathbf{k})$ to $\hat{H}(\mathbf{k})+\delta \hat{h}$, for small enough $\delta \hat{h}$ we have: 
\begin{align}
G^{-1}(\widetilde{\mathbf{k}})\rightarrow G^{-1}(\widetilde{\mathbf{k}}) + \delta \hat{h} \\ \nonumber
G(\widetilde{\mathbf{k}})\rightarrow G(\widetilde{\mathbf{k}}) -G(\widetilde{\mathbf{k}}) \delta \hat{h} G(\widetilde{\mathbf{k}})
\end{align}
From here on for simplicity we drop $\widetilde{\mathbf{k}}$ from our expressions. Applying the identities above we find the change in $\Pi_{a,b}(\mathbf{q},\omega)$ :

\begin{align}
\delta \Pi_{a,b}(\mathbf{q},\omega) &= e^2\frac{q}{6}\sum_{a,b,c}\varepsilon_{a,b,c}\int_{-\infty}^{\infty}\frac{d\omega}{2\pi}\int_{BZ}\frac{d^3k}{(2\pi)^3}Tr\Big\{\big[  \delta \hat{h} G \partial_{k_a} G^{-1} G \partial_{k_b} G^{-1} G \partial_{k_c} G^{-1} G \big]+\big[ \delta \hat{h} \partial_{k_a}(G \partial_{k_b} G^{-1}G \partial_{k_c} G^{-1}G) \big] \Big\} \\ \nonumber
&=e^2\frac{q}{6}\sum_{a,b,c}\varepsilon_{a,b,c}\int_{-\infty}^{\infty}\frac{d\omega}{2\pi}\int_{BZ}\frac{d^3k}{(2\pi)^3}Tr\Big\{\big[  \delta \hat{h} \partial_{k_a} G \partial_{k_b} G^{-1} \partial_{k_c}G \big]-\big[ \delta \hat{h} \partial_{k_a}(\partial_{k_b} G \partial_{k_c} G^{-1}G) \big] \Big\} \\ \nonumber
&=e^2\frac{q}{6}\sum_{a,b,c}\varepsilon_{a,b,c}\int_{-\infty}^{\infty}\frac{d\omega}{2\pi}\int_{BZ}\frac{d^3k}{(2\pi)^3}Tr\Big\{\big[  \delta \hat{h} \partial_{k_a} G \partial_{k_b} G^{-1} \partial_{k_c}G \big]-\big[ \delta \hat{h} \partial_{k_b} G \partial_{k_c} G^{-1}\partial_{k_a}G \big] \Big\}  \\ \nonumber
&=e^2\frac{q}{6}\sum_{a,b,c}\varepsilon_{a,b,c}\int_{-\infty}^{\infty}\frac{d\omega}{2\pi}\int_{BZ}\frac{d^3k}{(2\pi)^3}Tr\Big\{\big[  \delta \hat{h} \partial_{k_a} G \partial_{k_b} G^{-1} \partial_{k_c}G \big]-\big[ \delta \hat{h} \partial_{k_a} G \partial_{k_b} G^{-1} \partial_{k_c}G \big] \Big\} = 0
\end{align}
Where we have used cyclic properties of the trace, partial integrating and also the fact that any symmetric term inside the trace vanishes since the total answer is antisymmetric. So we established that $ \Pi_{a,b}(\mathbf{q},\omega)$ is a constant, to show that it's zero note that, if all greens functions have finite imaginary parts in their poles ( as they do in the disordered case ) , then the momentum integral includes no singularities and is therefore analytic. This means that we can continuously deform our hamiltonian into a constant  and force $ \Pi_{a,b}(\mathbf{q},\omega)$ to vanish but since we already proved that $ \Pi_{a,b}(\mathbf{q},\omega)$ is a constant under continuous deformations of the hamiltonian it follows that $ \Pi_{a,b}(\mathbf{q},\omega)$ has to be zero everywhere ( as long as there are no real poles ). Note that in the clean case this proof doesn't go through since green function's poles are  therefore real and the integrals are not analytic.

\end{widetext}

\end{document}